\documentclass[pra,twocolumn,nofootinbib]{revtex4}
\usepackage{amssymb,amsmath}
\usepackage{graphics}
\usepackage{rotating}
\begin{document}

\title{Angular distributions of scattered excited muonic hydrogen
atoms}
\thanks{This work was partially supported by Russian Foundation
 for Basic Research, grant No. 03-02-16616.}

\author{V.\,N. Pomerantsev, V.\,P. Popov}
\affiliation{Institute of Nuclear Physics, Moscow State
University\\}

\begin{abstract}
 Differential cross sections of the Coulomb deexcitation in the collisions of
 excited muonic hydrogen with the hydrogen atom have been studied for
the first time.  In the framework of the fully
quantum-mechanical close-coupling approach both the differential
cross sections for the $nl \to n'l'$ transitions and
$l$-averaged differential cross sections have been calculated
for exotic atom in the initial states with the principle quantum
number $n=2 - 6$ at relative motion energies $E_{\rm
{cm}}=0.01 - 15$~eV and at scattering angles $\theta_{\rm
{cm}}=0 - 180^{\circ}$. The vacuum polarization shifts of the $ns$-states are
taken into account. The calculated in the same approach
differential cross sections of the elastic and Stark scattering
are also presented. The main features of the calculated
differential cross sections are discussed and a strong anisotropy of 
cross sections for the Coulomb deexcitation is predicted. 
\end{abstract}


\maketitle

{\bf Introduction.} Exotic hydrogen-like atoms are formed
in excited states, when heavy negative particles ($\mu^-$,
$\pi^-$, etc.) are slowed down and captured in hydrogen
media. The following atomic cascade of collisional and
radiative transitions proceeds via many intermediate states
up to nuclear absorption or transition to the ground state
occurs. Since the experimental data are mainly available
for the last stage of this atomic cascade, the reliable
knowledge of the total and differential cross sections (DCS) of
the collisional processes during the cascade is needed for
the realistic analysis of these data.

 In particular, the collisional processes
 \begin{equation} \label{eq1}
(\mu^- p)_{nl} + H_{1s}\rightarrow(\mu^- p)_{n'l'} +H_{1s}
\end{equation}
of the elastic scattering ($n'\!=\!n,\,l'\!=\!l$), Stark
transitions ($n'\!=\!n,\,l'\!\neq \!l$), and Coulomb
deexcitation ($n'\!< \!n$)  essentially change the energy-
and $n l$-distributions of exotic atoms. It is especially
important from the view of the precise experiments at the
Paul Scherrer Institute with muonic~\cite{1} and
pionic~\cite{2} hydrogen atoms. These experiments are aimed
at the extraction of the root mean squared proton charge
radius with the relative accuracy of $10^{-3}$ from the
Lamb shift experiment~\cite{1} and the determination of the
$\pi N$ scattering lengths with the accuracy better than
$1$\% by extracting the shift and width of $1s$ state due
to the strong interaction in pionic hydrogen~\cite{2}. The
proper analysis of these experiments requires a reliable
theoretical treatment of related cascade processes.

The Coulomb deexcitation (CD) plays an important role in
the kinetic energy distribution of the exotic atoms. In
particular, the energy distribution of $(\mu^- p)$- and
$(\pi^- p)$- atoms during the radiative transitions
$np\rightarrow 1s$ have significant high-energy components
resulting from the preceding Coulomb deexcitation
transitions. Before the recent paper~\cite{3} the only
$\Delta n=1$ transitions are assumed to be important in the
Coulomb transitions at low $n$. Moreover, it is also
suggested in the cascade calculations~\cite{4} that the CD
process results in the isotropic angular distribution.

The main goal of this paper is to introduce the first
theoretical study of the Coulomb deexcitation differential cross
sections in the muonic atom -- hydrogen atom collisions. In
particular, we are interested in the main features of the $n$
and $E$ dependences of these cross sections. Our new results,
concerning the DCS of elastic and Stark
scattering are also presented.

 The first theoretical study of DCS for the  elastic and Stark scattering
  in the collisions of the
excited exotic atom from atomic hydrogen has been performed
within the quantum-mechanical adiabatic approach~\cite{5}.
Later, these cross sections were also calculated in the
framework of the close-coupling model and semiclassical
approximation~\cite{6} where the Coulomb interaction of the
exotic atom with the hydrogen atom field is modeled by the
screening dipole approximation. This approximation becomes 
invalid at low collisional energies below some value
$E^*$, which depends on the principal quantum number $n$
(see \cite{7}). Thus, the latter approach as well as
various modifications of the semiclassical model~
\cite{8,9} can result in uncontrolled errors in the
low-energy region where only a few partial waves are
important.

 In the present paper the DCS of the
processes (1) are studied in the framework of the more
accurate close coupling (CC) approach. The approach has
been developed earlier~\cite{10} by the authors to describe
the elastic scattering and Stark transitions in the exotic
atom -- hydrogen molecule collisions and employed
recently~\cite{3} to give a unified treatment of the
processes (1). In the framework of the CC approach the
first fully quantum-mechanical calculations of the total
cross sections of the  Coulomb deexcitation both for
muonic~\cite{3} and pionic~\cite{11} hydrogen atoms have
been performed.

{\bf Approach.} In the framework of the close-coupling
approach the total wave function of the four-body system is
expanded in terms of the basis states with the conserving
quantum numbers of the total angular momentum $JM$ and
parity $\pi = (-1)^{l +L}$. The basis states are chosen as
tensor products of the corresponding wave functions of the
free exotic and hydrogen atoms and the angular wave
function of their relative motion. This expansion results
in the set of the coupled differential equations. In
contrast with paper~\cite{6}, in the present approach the
interaction potential matrix for the exact four-body
Coulomb interaction of the colliding atoms is calculated
analytically. Moreover, as it is shown in paper~\cite{3}
the approximation similar to "dipole approximation" results
in an improper description of the CD process, especially
for the low-lying states of the exotic atom.

At fixed $E_{cm}$ (energy of collision in the center of
mass system) and given $J$ and $\pi$ the set of the coupled
equations are solved numerically by the Numerov method with
the standing-wave boundary conditions involving the real
and symmetrical $K$-matrix. The corresponding $T$-matrix
can be obtained from the $K$ matrix using the matrix
equation $T~=~2iK~(I~-~i~K)~^{-1}$. In the present study,
as distinct from paper~\cite{3}, we take into account the
energy shifts of the $ns$ states due to the vacuum
polarization which are very important especially at the low
kinetic energies comparable with the energy shift and for
the lower states of the exotic atom. The $2s - 2p$ energy
splitting $\Delta \varepsilon _{2s-2p}$ for the muonic
hydrogen atom is equal to 0.206~eV and this splitting
decreases approximately as $n^{-3}$ with $n$ increasing.

The formalism has been described in more details
in~\cite{3}. Here we give only the formulas for the DCS.
The differential cross sections for the transition from the
initial state $(n l)$ to the final state $(n'l')$ are
defined as

\begin{equation} \label{eq2}
\frac{d\sigma_{nl \rightarrow n'l'}}{d\Omega}
=\frac{1}{2l+1}\frac{k_f}{k_i}\sum_{m m'}|f_{nlm\rightarrow
n'l'm'}(k_i, k_f; \Omega )|^2,
\end{equation}
where the scattering amplitude for the transition $nlm
\rightarrow n'l'm'$ is given by
\begin{align}
 f_{nlm\rightarrow n'l'm'}(k_i, k_f; \Omega ) =  \frac{2\pi i}{\sqrt{k_{i}k_{f}}}
   \sum_{JLL'\lambda'}i^{L'\!-\!L} \nonumber\\
   Y_{L0}^{*}(0)
   \langle lmL0|J m\rangle T^J_{nlL\rightarrow n'l'L'}
 \langle l'm'L'\lambda'|J m\rangle Y_{L'\lambda'}(\Omega ).
  \label{eq3}
\end{align}
 Here, $k_i$ and $k_f$ are the cms relative momenta
in the initial and final channels, correspondingly;
$\Omega\equiv \theta, \varphi$, where $\theta$ is the cms
scattering angle, and $T^J_{i \rightarrow f}$ is the
transition matrix in the total angular momentum
representation. The indices of the entrance channel and the
target electron state are omitted for brevity.

In order to illustrate the most general features of DCS, it
is also useful to introduce the cross sections averaged
over the initial orbital angular momentum $l$ of the exotic
atom. So, the following $l$-averaged angular distributions
are also discussed in the present study:\\ for elastic
scattering
\begin{equation} \label{eq4}
\frac{d\sigma^{el}_{n}}{d\Omega} =\frac{1}{n^2}\sum_{l} (2l+1)
\frac{d\sigma_{nl \rightarrow nl}}{d\Omega},
\end{equation}
for Stark transitions
\begin{equation} \label{eq5}
\frac{d\sigma^{St}_{n}}{d\Omega} =\frac{1}{n^2}
\sum_{l, l'}(1- \delta_{ll'})(2l+1)
\frac{d\sigma_{nl \rightarrow nl'}}{d\Omega},
\end{equation}
and, finally, for CD process
\begin{equation} \label{eq6}
\frac{d\sigma^{CD}_{n\rightarrow n'}}{d\Omega} =\frac{1}{n^2}\sum_{l, l'}
(2l+1)\frac{d\sigma_{nl \rightarrow n'l'}}{d\Omega},\qquad n'< n.
\end{equation}
Hereafter, the atomic units
will be used throughout the paper and the collision energy will
be referred to the states with $l\neq 0$ in the entrance
channel, which are assumed to be degenerated.

{\bf Results.} The numerical calculations of the
DCS for the collisional processes
(1) have been done for $(\mu p)_n$ atoms with the initial
principal quantum number values $n=~2-6$ at the relative
motion energies $E_{\rm {cm}}=~0.01~-~15$~eV and at all the
scattering angles $\theta_{\rm {cm}}$ from zero up to
$180^{\circ}$. All the exotic-atom states corresponding to
the open channels have been included in the close-coupling
calculations. Some of our results are present here in
Figs.~1~-~9 both for the cross sections of the separate
$nl\rightarrow n'l'$ transitions and for the $l$-averaged
ones.

\begin{figure}[h]
\includegraphics[width=0.46\textwidth,keepaspectratio]{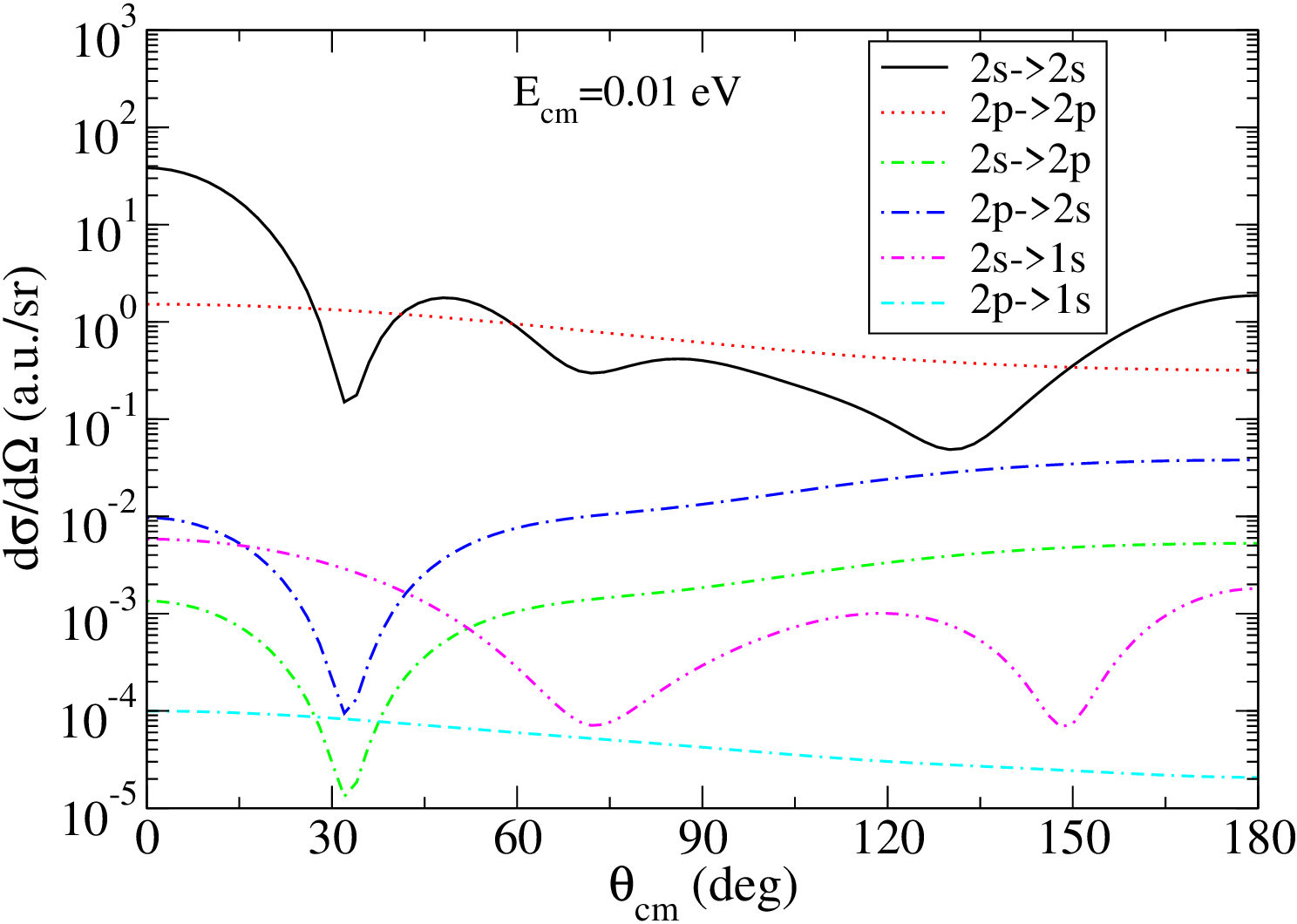}
     \caption{Differential $2l\rightarrow 2l'$ and $2l\rightarrow 1s$
     cross sections for $(\mu p)_{2l} + H$ collisions vs. cms
     scattering angle $\theta_{cm}$ at $E_{cm} = 0.01$~eV (referring to
     $2p$ state).}
 \label{fig1}
     \end{figure}

In Figs.~1 and  2 the DCS  for $2p\rightarrow 2p, 2s$
(elastic and Stark scattering) and $2p\rightarrow 1s$
(Coulomb deexcitation) transitions versus cms scattering
angle $\theta_{\rm {cm}}$ at energies $E_{\rm {cm}}~=~
0.01$~eV and $1$~eV (referring to the $2p$ threshold) are
shown. It is worthwhile noting that the relative motion
energy in the entrance channel for the $2s\rightarrow 2s,
2p$  and $2s\rightarrow 1s$ transitions increases due
to the Lamb shift $\Delta \varepsilon _{2s-2p}=
0.206$~eV in comparison with the scattering processes of the
muonic hydrogen atom in the $2p$ state.
As it is seen from Fig.~1, the angular distributions of the
elastic $2p\rightarrow 2p$ scattering and Coulomb
$2p\rightarrow 1s$ transition are similar and almost
isotropic, their shapes are  mainly defined by the
contributions of the  $S$-wave relative motion with a small
mixture of the $P$-wave.

\begin{figure}[h!]
    \includegraphics[width=0.46\textwidth,keepaspectratio]{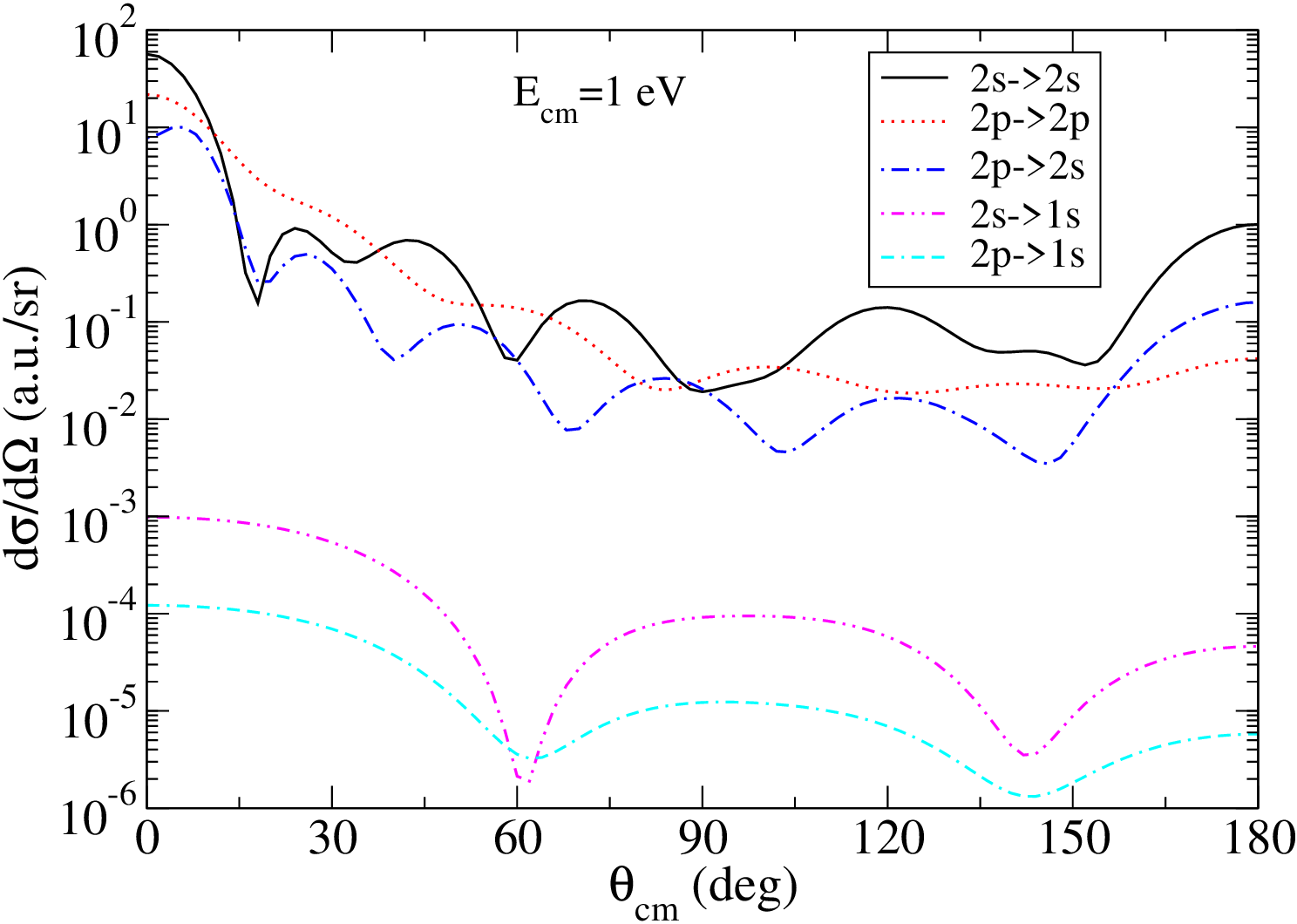}
     \caption{The same as in Fig.~1 but at $E_{\rm {cm}} = 1$~eV.}
 \label{fig2}
     \end{figure}

It is also seen that Coulomb deexcitation process for $2p
\rightarrow 1s$ transition is more than four order of the
magnitude suppressed relatively to the $2p\rightarrow 2p$
elastic scattering one and more or about two order of the
magnitude as compared with both the Stark $2p\rightarrow
2s$ and Coulomb $2s\rightarrow 1s$ transitions,
respectively.

The Stark transitions $2p\rightarrow 2s$ and $2s\rightarrow
2p$ are strongly inelastic at such a low energy (the
initial and final energies differ more than twenty times)
and that results in their quite unusual angular
distributions (see Fig.~1) with the maximum at zero
scattering angle and minimum at $\theta_{\rm {cm}} \sim
30^{\circ}$ and smooth increasing at the backward
hemisphere.

In the DCS of the elastic $2s \rightarrow 2s$ scattering
($E_{\rm {cm}}=0.216$~eV) we see the structures (due to
higher partial waves, involved in the scattering process)
as compared with the almost isotropic DCS for
$2p\rightarrow 2p$ transition. The comparison of the DCS of
the Stark transitions, presented in Figs. 1 and 2, shows a
significant effect of the Lamb shift on these cross
sections, which changes substantially depending on the
ratio value of the  $E_{\rm {cm}}/\Delta \varepsilon
_{2s-2p}$. The Stark DCS of the $2p\!\to \!2s$ and
$2s\!\to\! 2p$ transitions at low kinetic energy
comparable with the Lamb shift value have quite different
behavior in comparison with the elastic ones and reveal a
strong suppression in the forward hemisphere by more than
three order of the magnitude. When the energy of the
collision is much larger than $\Delta \varepsilon _{2s-2p}$
we observe the usual picture~\cite{5,6} of the DCS for
Stark transitions with a strong forward peak and a set of
maxima and minima (see Fig.~2).

Coulomb deexcitation process occurs at essentially smaller
distances than the elastic and Stark processes. So, the
number of partial waves involved in the CD process, as a
rule, is much smaller, as a strong centrifugal barrier
prevents the colliding atoms from penetrating in the
interaction region corresponding to the process. With
decreasing the value $n$ the number of the partial waves
contributing to the CD process (at the fixed energy) also
decreases.

In particular, the DCS of the CD process in case of $n=2$
have a quite simple angular dependence due to a few partial
waves involved in the process( see Fig. 1 and 2). The
angular dependences of the DCS for $2s\rightarrow ~1s$ and
$2p\rightarrow ~1s$ Coulomb transitions are mainly
determined by the contributions of the lowest partial waves
of the relative motion at all energies under consideration.
The differential cross sections of these Coulomb
transitions as it is seen in Fig.2 have a similar angular
dependence which shape is slowly changed enhancing the
forward hemisphere scattering with energy increasing.

It is well-known that in the muonic hydrogen atoms, the
$2s$-state plays a particular role due to $2s$ Lamb shift
and has no analog in the other exotic atoms in which the
strong interaction leads to a large rate of the nuclear
absorption from this state. In particular, a new knowledge
about the collisional quenching of $2s$ state at
collisional energy near or below the $2p$ threshold is of
special interest.

As it shown in Fig. 1, the DCS for the $2p\!\to\!1s$
transition is strongly suppressed in comparison with the
main $2s\rightarrow ~1s$ transition about two order of the
magnitude and this suppression  is also observed at  higher
kinetic energy (see also Fig. 2).  Hence the $2s\rightarrow
~1s$ transition determines the CD $2\rightarrow 1$
transition at all kinetic energies and it is quite probable
that the observed collisional quenching of the metastable
$2s$ state of the muonic atom and the high energy component of
muonic hydrogen in $1s$ state can be explained by the
direct Coulomb deexcitation process.

 \begin{figure}[h!]
 \includegraphics[width=0.46\textwidth,keepaspectratio]
 {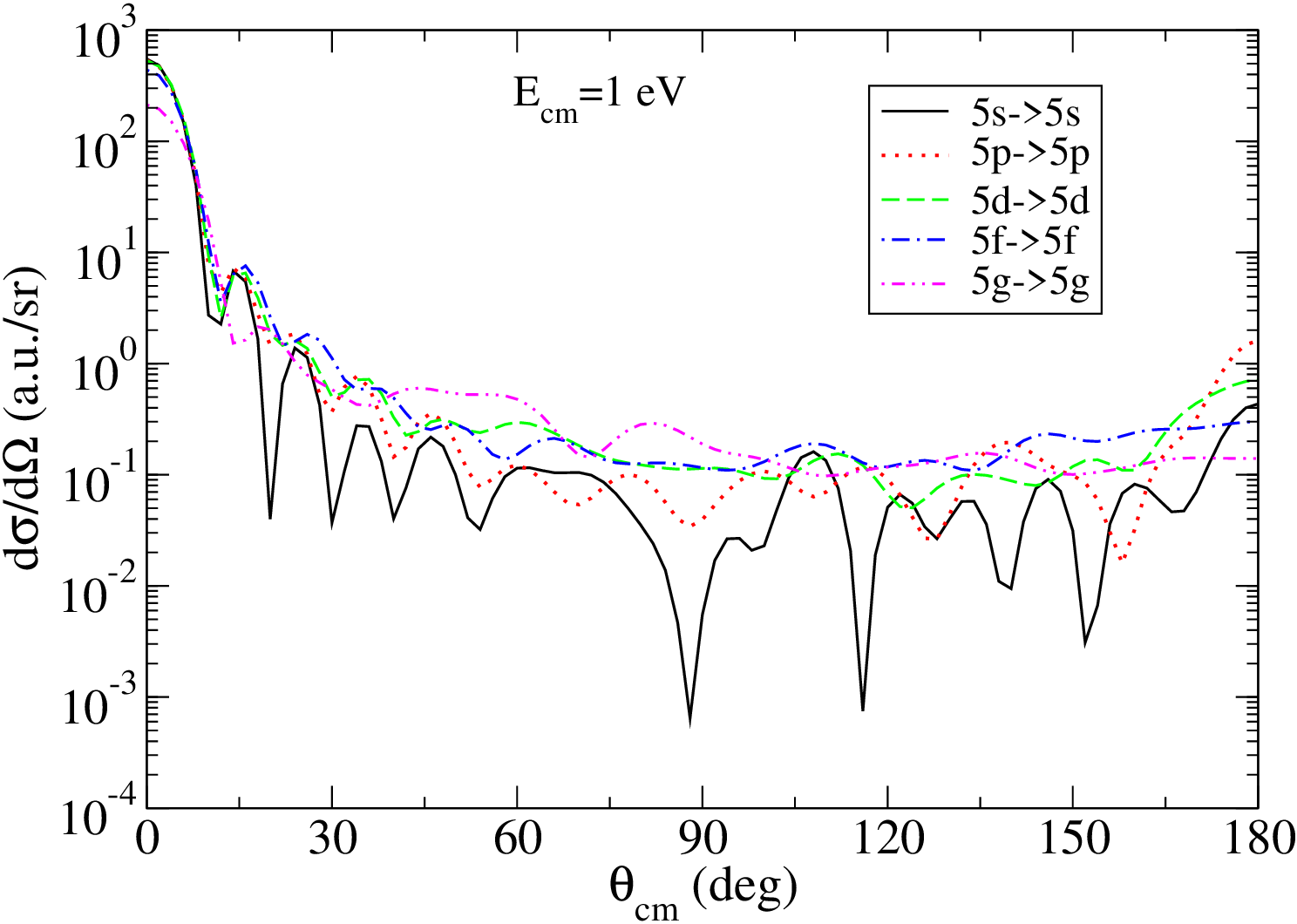}
  \caption{Differential elastic ($l=l'$)
     cross sections for $(\mu p)_{5l} + H$ collisions vs. cms
     scattering angle $\theta_{\rm cm}$ at $E_{\rm {cm}} = 1$~eV.}
 \label{fig3}
     \end{figure}

The typical angular distributions for the
individual $nl\rightarrow nl'$ transitions for $n=5$ are
shown in Fig.~3 and Fig.~4 for the elastic scattering and
Stark transitions, respectively. It is well
known~\cite{5,6} that DCS of these processes  are similar
 to the diffraction scattering (at the collisional energies
 more or about $1$~eV) with a
strong forward peak which is enhanced with increasing
energy and a set of maxima and minima. While the elastic
cross sections always have a strong peak at $\theta_{\rm
{cm}}=0$,
 the first maximum position in the Stark DCS depends on
 the $\Delta l=|l-l'|$
 value. In particular, for $\Delta l=1$ this maximum is
at finite scattering angles as it also remarked in~\cite{6}. 
According to our calculations,
the sharpest variations in DCS are always observed in the
$ns\rightarrow n's$ and $ns\rightarrow n'p$ transitions
(see also Fig. 7 for DCS of the CD process).

  \begin{figure}[h]
  \includegraphics[width=0.46\textwidth,keepaspectratio]{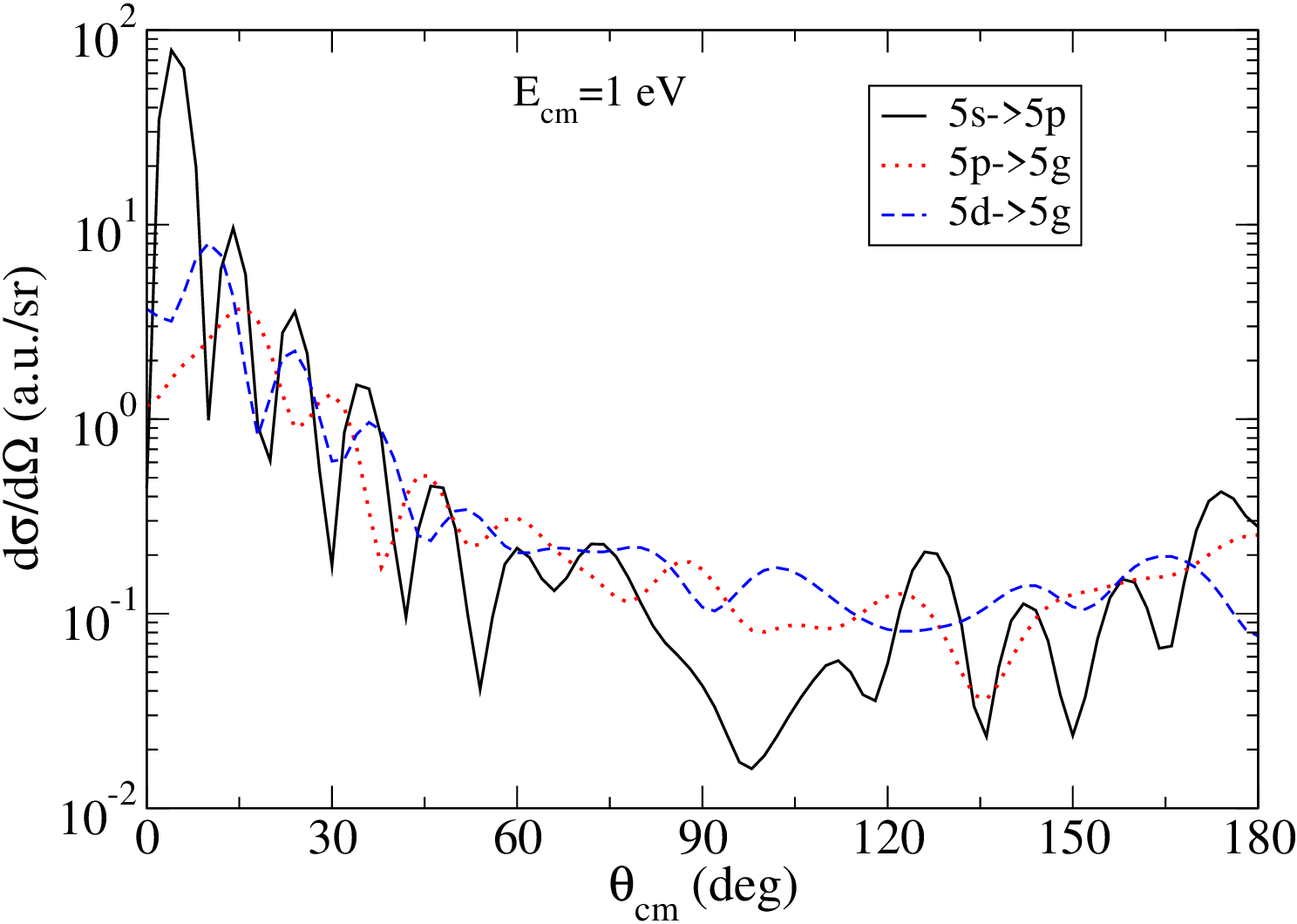}
  \caption{Differential Stark cross sections for  $5l\rightarrow 5l'$
  transitions with $\Delta l = 1 - 3$ in $(\mu p) + H$ collisions at 
  $E_{\rm cm} = 1$~eV.}
 \label{fig4}
     \end{figure}

The first and next peaks in the forward hemisphere for the
elastic scattering (see Fig.~3) and Stark transitions (see
Fig.~4) have a tendency to be less pronounced  and the
angular distribution becomes smoother with increasing $l$ and $\Delta l $
is increased. According to our calculations the shape of
the peaks in the forward hemisphere is sharper with $n$
increasing at the fixed collisional energy.

 \begin{figure}[h]
 \includegraphics[width=0.46\textwidth,keepaspectratio]{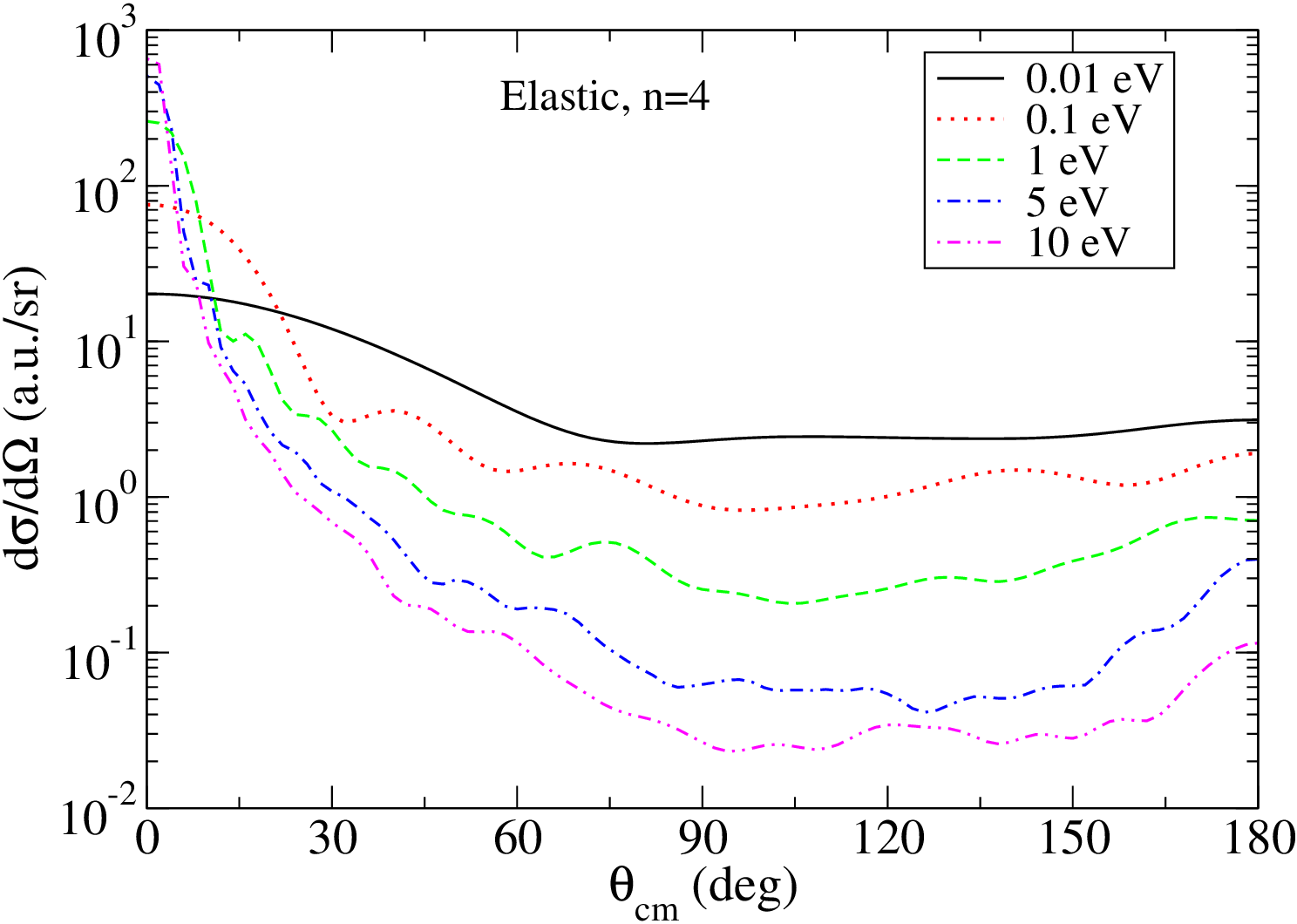}
  \caption{ The $l$-average differential elastic cross sections
  for $(\mu p)_{n=4} + H$ collisions at
  different energies.}
 \label{fig5}
  \end{figure}

 The dependence of the $l$-averaged DCS for the elastic
 scattering on the collisional energy is shown in Fig.~5
 for $n=4$. While the
DCS for the individual elastic $nl\rightarrow nl$
transitions (the same is valid for the Stark $nl\rightarrow
nl'$ transitions) reveal the complicated structure, the
$l$-averaged cross sections smooth out many details and
allow to study the most general features of the process.
Figure 5 shows, that at low energies ($\lesssim 1$~eV) the
DCS can be approximately considered as constant for the
simple estimation at scattering angles $\theta_{cm}$ more
or about $75^{\circ}$. However, at higher
energies the appreciable enhancing of the backward
scattering is observed.
   \begin{figure}[h]
  \includegraphics[width=0.46\textwidth,keepaspectratio]{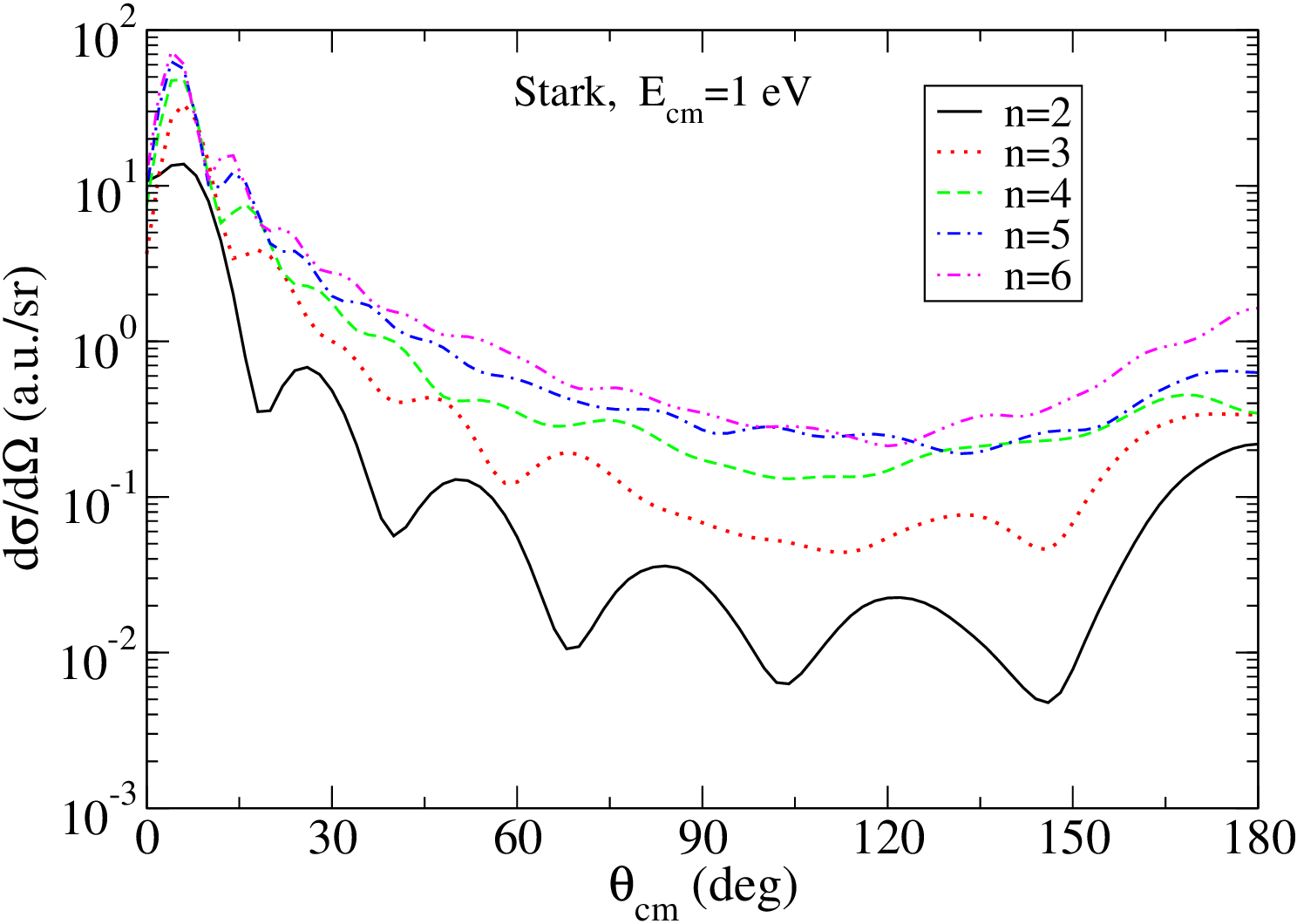}
  \caption{ The $l$-average differential Stark cross sections 
  for  $(\mu p)_{n} + H$ collisions ($n=2 - 6$) at $E_{\rm {cm}} = 1$~eV.}
   \label{fig6}
  \end{figure}

In Fig.~6 we show the $l$-averaged Stark DCS at $E_{\rm
cm}=1$~eV for different values of $n$. Here one can see
that with the increase of $n$ the first forward peak also becomes
sharper and narrower (as well as for the elastic scattering)
but remains to be at the finite values of scattering angle.
The height of this peak depends on $n$ not so strong as the
diffraction maximum in elastic scattering (cf. Figs.~5 and
6). The diffraction structure of minima and maxima becomes
less pronounced with increasing $n$. As a whole our results
for the elastic and Stark DCS are in a qualitative
agreement with the previous calculations~\cite{5,6}.

Now we are coming to the discussion of the typical  angular
distributions for the CD process. As far as we know, the
calculations of these DCS have not been reported until now
and in the cascade calculations the angular distributions
of the CD process are presumed to be isotropic. The
calculated DCS for individual $nl\rightarrow n'l'$
transitions with $\Delta n=1$ and 2 at relative motion
energy  $E_{\rm {cm}} = 1$~eV are shown in Figs. 7 and 8,
respectively. In Fig.~9 the $l$-averaged DCS for the
$6\rightarrow 5$ transition at different values of the
relative motion energy from 0.01 up to 15~eV are presented.

Our study reveals the following main features of the CD angular distributions.
The angular distributions both of the individual and $l$-averaged 
cross sections (excluding very low energies) 
 are far from isotropic: as a whole
    the scattering  at $\theta_{cm} \lesssim 60^{\circ}$ and 
    $\theta_{cm} > 120^{\circ}$ is noticeably enhanced.

 \begin{figure}[h!]
   \includegraphics[width=0.46\textwidth,keepaspectratio]{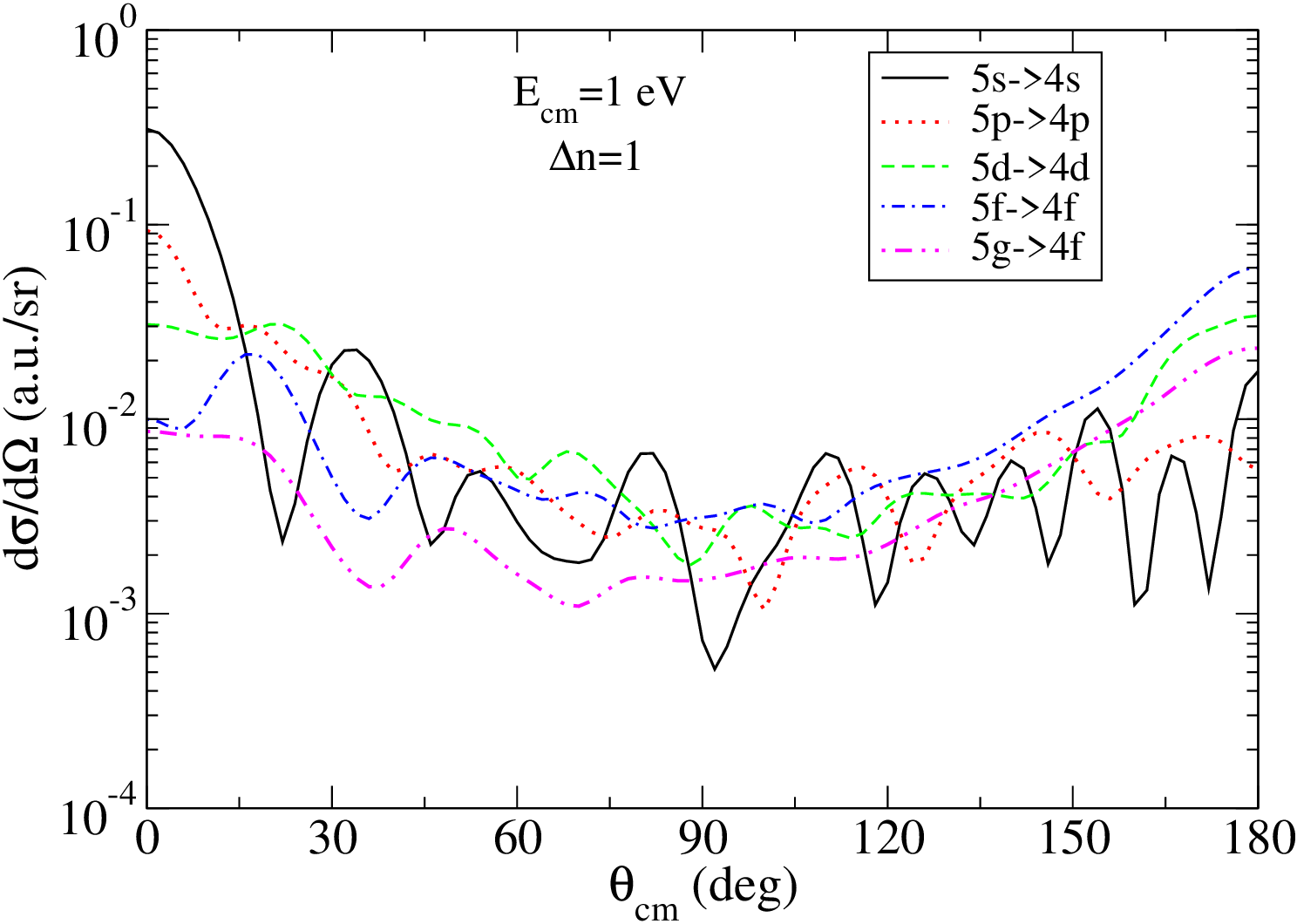}
  \caption{Differential CD cross sections for the
  individual transitions with $\Delta n=1$ for $n=5$  at  $E_{\rm {cm}} = 1$~eV.}
  \label{fig7}
   \end{figure}

 \begin{figure}[h!]
 \includegraphics[width=0.46\textwidth,keepaspectratio]{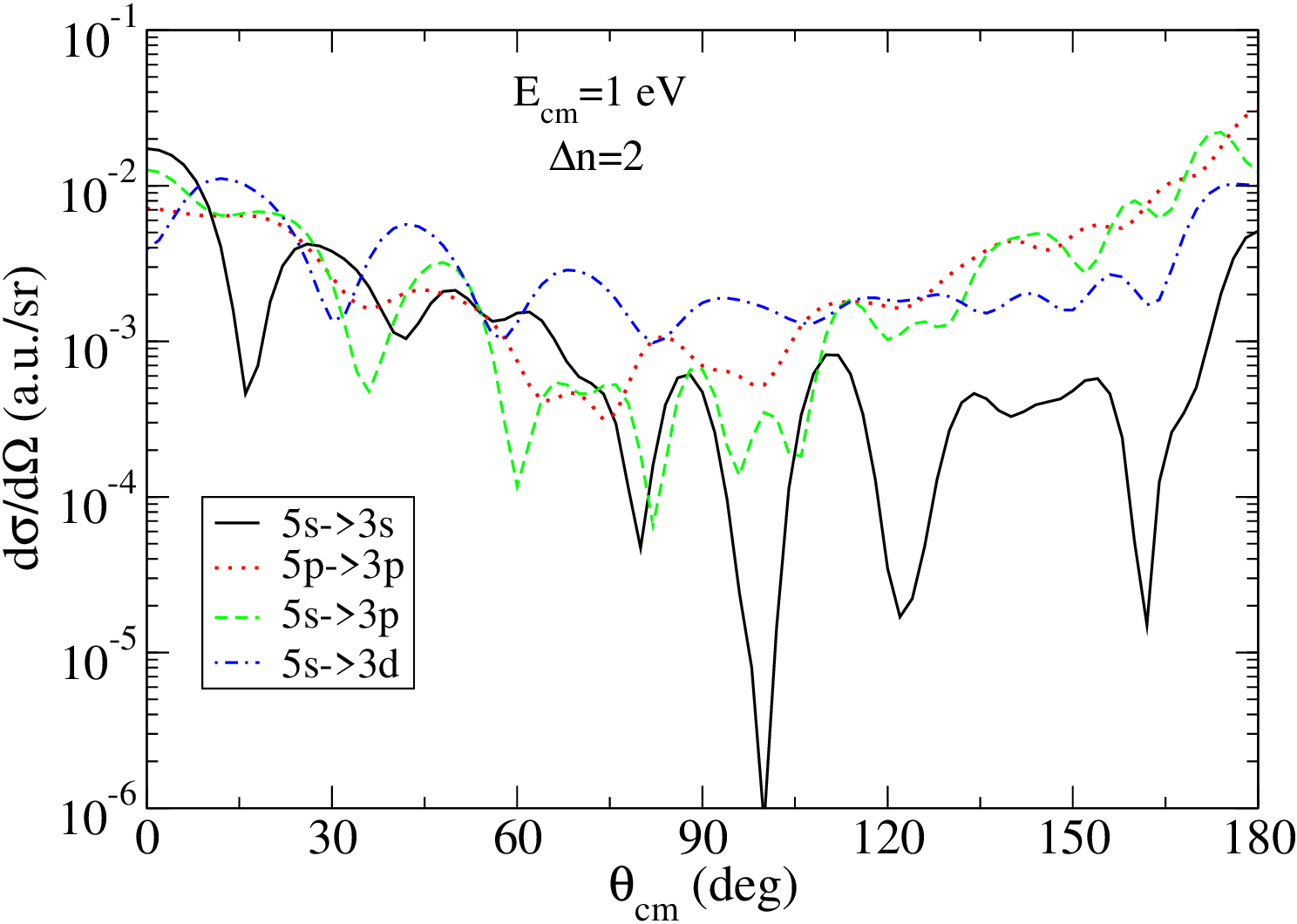}
\caption{The same as in Fig.~7 but for the $\Delta n=2$ transitions.}
 \label{fig8}
 \end{figure}

   \begin{figure}[h!]
  \includegraphics[width=0.46\textwidth,keepaspectratio]{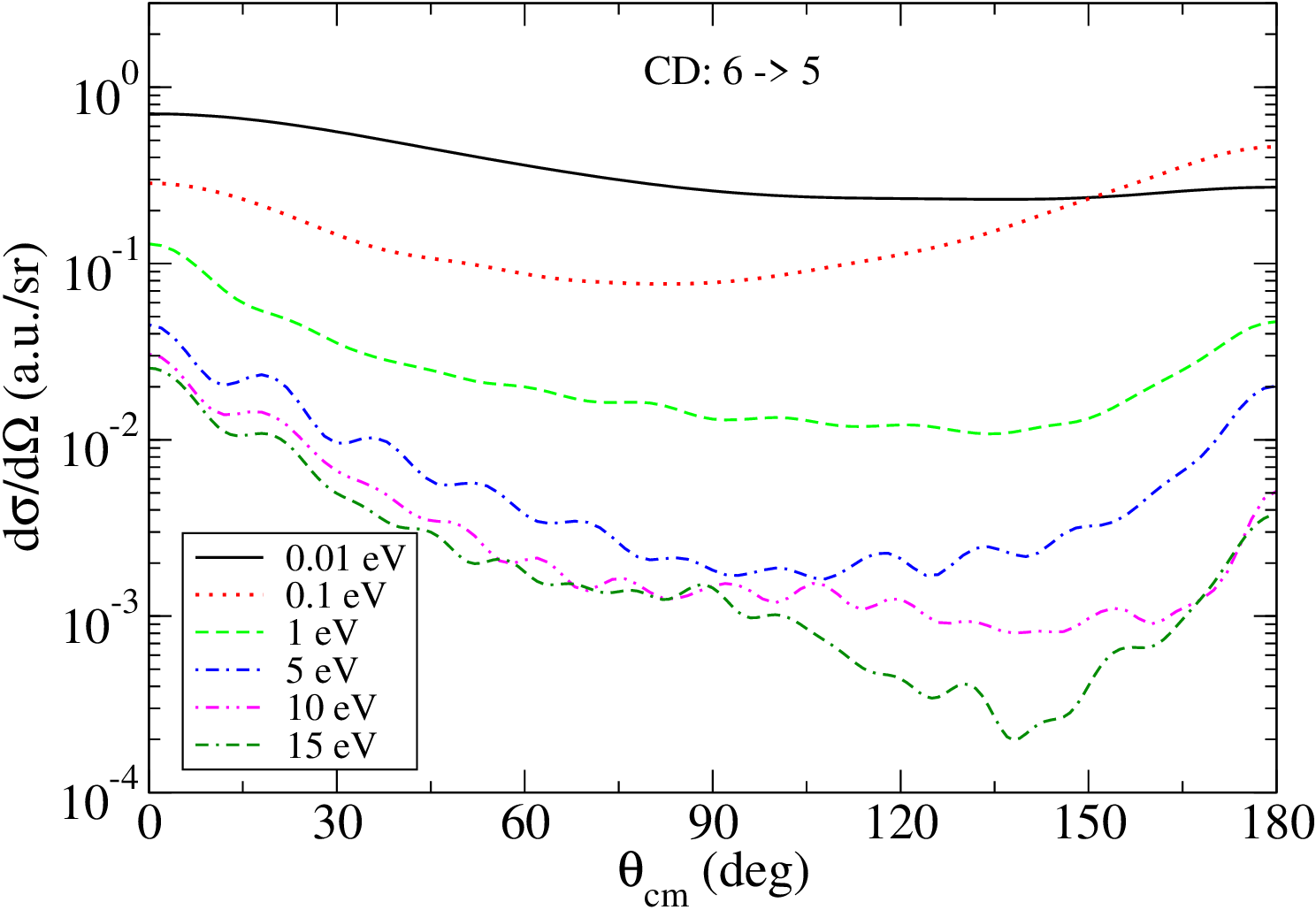}
  \caption{ The $l$-average CD differential cross sections
  for transition $6 \to 5$ at different energies.}
   \label{fig9}
 \end{figure}
 
 The DCS for  $ns\rightarrow n's$ transitions (see Figs.~7, 8) have (as in case
 of the elastic scattering) a more pronounced diffraction structure with sharp
 maxima and minima and a strong peak at zero angle as compared with the 
 smoother angular dependence for the other CD  transitions. This behaviour can
 be simply explained by the conditions $L=L'=J$ (for $ns\to n's$ transitions)
 which strongly reduce the number of terms in the amplitude (3) in contrast
 with the  other transitions.

 The increase of kinetic energy enhances asymmetry in
the  angular dependence of the $l$-averaged DCS and
decreases the role of the backward scattering (see Fig.~9).

{\bf Summary.} The fully quantum-mechanical CC approach has been applied for
the calculations of the elastic scattering, Stark transition and Coulomb
deexcitation DCS in a self-consistent manner and the detailed analysis of the
obtained results has been performed. For the first time the DCS of the CD
process have been calculated for the values of the principal quantum number and
kinetic energy relevant for kinetics of the atomic cascade. The first results
for the direct collisional quenching of the $2s$-state due to CD process were
also obtained. The present study reveals the new knowledge about the CD process
and is very important for the reliable analysis of the $K$ $X$-ray yields and
high energy component in the kinetic energy distribution of muonic hydrogen
atoms. We hope that our study allows to remove some uncertainties inherent in
the previous cascade calculations, which resulted from the treatment of
collisions, especially Coulomb deexcitation, involving different and not always
self-consistent approximations.

 We are grateful to Prof. G. Korenman for fruitful discussions.

\end{document}